\documentclass[aps,prb,10pt,twocolumn,superscriptaddress,floatfix,amsmath,amssymb]{revtex4-1}
\usepackage{hyperref}
\usepackage[usenames,dvipsnames]{color}
\usepackage[english]{babel}
\usepackage{graphicx}
\usepackage{epstopdf}
\usepackage{lineno}
\usepackage{tabularx}
\usepackage{blindtext}
\usepackage{nicefrac}

\begin{document}
\title{Charge order and antiferromagnetism in the extended Hubbard model}

\author{Joseph Paki} 
\affiliation{Department of Physics, University of Michigan, Ann Arbor, MI 48109, USA}
\author{Hanna Terletska}
\affiliation{Department of Physics and Astronomy, Computational Science Program, Middle Tennessee State University, Murfreesboro, TN 37132, USA}
\author{Sergei Iskakov}
\affiliation{Department of Physics, University of Michigan, Ann Arbor, MI 48109, USA}
\author{Emanuel Gull}
\affiliation{Department of Physics, University of Michigan, Ann Arbor, MI 48109, USA}
\affiliation{Center for Computational Quantum Physics, Flatiron Institute, New York, NY 10010, USA}

\date{\today}
\begin{abstract}
We study the extended Hubbard model on a two-dimensional half-filled square lattice using the dynamical cluster approximation.  We present results on the phase boundaries between the paramagnetic metallic (normal) state and the insulating antiferromagnetic state, as well as between the antiferromagnetic and charge order states.  We find hysteresis along the antiferromagnet/charge order and normal/charge order phase boundaries (at larger values of the on-site interaction), indicating first order phase transitions.  We show that nearest neighbor interactions lower the critical temperature for the antiferromagnetic phase. We also present results for the effect of nearest neighbor interactions on the antiferromagnetic phase boundary and for the evolution of spectral functions and energetics across the phase transitions.
\end{abstract}
\maketitle

%\linenumbers

\section{Introduction}
Strongly correlated electron systems with many degrees of freedom often exhibit complex phase diagrams with a wide range of phases.\cite{Dagotto_science} Competing interactions may lead to symmetry breaking charge, spin, superconducting, or orbital ordered states. Of special interest are systems that display several ordered states in close proximity, such as charge order and magnetism, as other types of orders (such as superconductivity) often occur near their respective phase boundaries. 

Compounds that exhibit both charge ordered (CO) and antiferromagnetic (AFM) phases are ubiquitous in nature.\cite{Imada98, cuprates,Dagotto_science} Examples include the $d$-electron material La$_{1-x}$Sr$_x$FeO$_3$,\cite{Battle90} the doped nickelate La$_{2-x}$Sr$_x$NiO$_4$,\cite{Tranquada95} the layered manganite La$_{0.5}$Sr$_{1.5}$MnO$_{4}$,\cite{Sternlieb96} the cobalt oxides\cite{cobalts}, the doped iridate \cite{Iridate_Luyan}, the layered ruthenate \cite{Ruthenate} and the layered cuprates La$_{2-x}$Sr$_x$CuO$_4$ and La$_{2-x}$Ba$_x$CuO$_4$ at $1/8$ doping.\cite{Bednorz86}
Organic salts, including the one-dimensional (TMTTF)$_2$SbF$_6$\cite{Jerome91,Yu04,Nad06,Matsunaga13} and two-dimensional quarter filled compounds\cite{Seo00,Dressel04,McKenzie} similarly show coexisting AFM and CO. Several of these materials are also superconducting.
Understanding the phase diagram in these materials requires a detailed analysis of the competition between these two types of ordering.

Fermion model systems aim to capture the main aspects of these materials while abstracting the complexity of the underlying electronic structure problem. The most simple of these models is the Hubbard model in two dimensions, which has become the archetype of strongly correlated electron systems.\cite{LeBlanc15} The model approximates the band structure by a single band with nearest-neighbor hopping $t$ and on-site interaction $U$. It is known to have both strong short-ranged AFM correlations\cite{Toschi15,Gull09_8site,Gull10_clustercompare,Wu17} and a charge ordered ground state at $\nicefrac{1}{8}$ doping.\cite{Zheng17}

While CO in the two-dimensional (2D) Hubbard model on a square lattice is rather fragile, the extended Hubbard model promotes CO by the explicit addition of a repulsive nearest-neighbor interaction term $V$. The non-local interactions have been found to be sizable in a number of low-dimensional materials, resulting in CO as well as strong screening effects.\cite{Wehling,silke,Solyom, Schuler,Tom2019} The inclusion of non-local inter-site interactions energetically favors breaking translational symmetry and generating checkerboard CO states with two electrons on one site, none on its nearest neighbors, and a repeating $(\pi,\pi)$ charge ordered pattern.\cite{Zhang1989} In contrast, a large on-site interaction $U$ will enhance AFM $(\pi,\pi)$ correlations. \cite{Fuchs2011} The interplay between inter-cite interaction $V$, local interaction $U$, temperature, and doping effects thereby generates the rich phase diagram of the model.

The extended Hubbard model has been of interest both as a proxy for the exploration of charge order caused by electron repulsion\cite{Zhang1989,hirsch,Potthoff,Wu_Tremblay,Wehling,silke,silke_Fock,Schuler,Ayral2013,Kapcia2017,Haule,Rosciszewski,Merino_qcp, Merino07,Gao,Amaricci,Davoudi06, Davoudi07,Davoudi08,vanLoon,Iskakov,Chauvin17,Terletska2017,Terletska2018} and as a model system for testing the effect of non-local interactions on electron correlations.\cite{Senechal13,Schuler18,Jiang18, Tremblay_sc_afm, Schuler_2019} In that context, it has been particularly valuable to illustrate the convergence of diagrammatic extensions of the dynamical mean field theory\cite{Rohringer18, Werner_2016} (DMFT) including the GW+DMFT\cite{Kotliar_GW,Ayral2013,silke_Fock} and the dual boson approximation.\cite{Rubtsov2012,vanLoon,vanLoon2,DB_nature,Stepanov2019} 

Real materials that exhibit CO in the vicinity of AFM are considerably more complex than the simple extended Hubbard model. Nevertheless, there is merit in identifying model systems and non-perturbative approximations in which those phases occur in close proximity, as simple competition effects such as the one between local and non-local interactions here can provide general guiding principles for understanding their overall behavior.

In this paper we study the interplay between CO and AFM in the 2D half-filled extended Hubbard model at non-zero temperature using the dynamical cluster approximation (DCA)\cite{Hettler98,Maier2005} on an $8$ site cluster. This non-perturbative numerical method allows the explicit inclusion of non-local interactions and correlations and treats charge and spin correlations on equal footing. Previous DCA work in the absence of AFM order showed the detailed finite temperature phase diagram at\cite{Terletska2017} and away from\cite{Terletska2018} half-filling, demonstrating that an increase of $V$ at fixed $U$ leads to a checkerboard pattern of electrons characterized by a staggered density. This phase appears below a critical temperature which strongly depends on $U$ and $V$. We also found that non-local interactions cause noticeable screening effects.  Our study extends this work by allowing for both AFM and CO. This allows us to explore finite temperature phase transitions between AFM and CO and illustrate the effect of non-local interactions on the AFM phase. 

The exact mathematical solution of the Hubbard model does not support AFM order at non-zero temperature, as long-ranged antiferromagnetic fluctuations will always destroy this order.\cite{Mermin66} However, many numerical methods such as the DCA operate on a finite system and thereby truncate the correlation length, either suppressing longer ranged fluctuations entirely or supplanting them by infinitely ranged order \cite{Jarrell_2001,Maier2005, Maier_clusters, Lichtenstein00,Senechal_cluster_methods,Senechal_sc_afm,Tremblay_afm}. In our study, we use this truncation to generate a system that exhibit both CO and AFM in order to study the generic interplay between those phases.

The remainder of this paper is organized as follows. In Sec.~\ref{sec:Methods}, we discuss the Hamiltonian, our approximation, and the numerical method we use. In Sec.~\ref{sec:Results} we present and discuss our results. We first show the phase diagrams in the space of $T-V$ 
and $V-U$.
We then explore the corresponding CO and AFM order parameters and their temperature $T$ and non-local interaction $V$ dependence, as well as the details of the phase boundaries (Sec.~ \ref{ssec:order_parameter}).
Finally, we discuss the energetics (Sec.~\ref{ssec: Energetics}) across the phase transitions, the hysteresis behavior (Sec. ~\ref{ssec:hysteresis}), and the evolution of spectral functions (Sec.~\ref{ssec:spectral_function}) across the phase boundaries. We present our conclusions in Sec.\ref{sec:Conclusions}.

\section{Model and Methods}\label{sec:Methods}
This work applies the methods developed in Ref.~\onlinecite{Terletska2017} and Ref.~\onlinecite{Fuchs2011} to study the formation and competition between AFM and CO phases in the half filled 2D extended Hubbard model on a square lattice. The following provides an overview of the formalism, and the interested reader is referred to Ref.~\onlinecite{Terletska2017} for further details.

The Hamiltonian for the extended Hubbard model on a 2D  square lattice is given by
\begin{align}
H&=-t\sum_{\langle ij\rangle,\sigma}\left ( c_{i\sigma}^{\dagger}c_{j\sigma}+c_{j\sigma}^{\dagger}c_{i\sigma}\right )+U\sum_{i}n_{i\uparrow}n_{i\downarrow}  \nonumber \\
&+\frac{V}{2}\sum_{\langle ij\rangle,\sigma\sigma'}n_{i\sigma}n_{j\sigma'}-\tilde{\mu}\sum_{i\sigma}n_{i\sigma},
\label{Hamiltonian}
\end{align}
where $t$ is the nearest-neighbor hopping amplitude, $U$ and $V$ represent the on-site and nearest neighbor Coulomb interactions,  and $\tilde{\mu}$ denotes the chemical potential. $c_{i\sigma}^{\dagger} (c_{i\sigma})$ is the creation (annihilation) operator for a particle with spin $\sigma$ on lattice site $i$, and the particle number operator for site $i$ is $n_{i\sigma}=c_{i\sigma}^{\dagger}c_{i\sigma}$. Throughout this paper we restrict our attention to the half filled system, which occurs at $\tilde{\mu}=\mu_\text{HF}=\frac{U}{2}+4V$ for the 2D square lattice. We use dimensionless units $U/t$, $V/t$, $\beta t$, and $\mu/t$, and set $t=1$.

We compute our results within the DCA \cite{Hettler98,Maier2005} to find approximate solutions for the lattice model. The DCA is a cluster extension of the Dynamical Mean Field Theory (DMFT)~\cite{Georges96} that approximates the infinite lattice problem by a finite size cluster coupled to a non-interacting bath. The coupling to the bath is adjusted self-consistently by coarse-graining the Brillouin zone into $N_c$ momentum space patches. The self-consistency condition requires that certain cluster quantities (such as the Green's function) match the corresponding coarse-grained lattice quantities.\cite{Maier2005} The scheme becomes exact in the limit where $N_c \rightarrow \infty$, and recovers DMFT for $N_c=1$. The method is able to describe short-ranged spatial correlations non-perturbatively (i.e. correlations on length scales smaller than $N_c$), but correlations outside the cluster are neglected.  The method is also capable of simulating ordered phases, as long as the symmetry breaking is commensurate with the cluster.\cite{Maier2005,Fuchs2011}  An important detail is that for the extended Hubbard model the DCA coarse-graining procedure renormalizes the nearest neighbor interaction $V$ as $\bar{V}=\sin(\pi/N_c)/(\pi/N_c) V$, as described by Ref.~\onlinecite{Arita04} and Ref.~\onlinecite{Wu_Tremblay}. In this paper we study systems with $N_c=8$.

We can bias the system towards an ordered phase by adding symmetry breaking terms to the Hamiltonian.\cite{Maier2005, Terletska2017,Fuchs2011}  These terms extend Eq.~\ref{Hamiltonian} by a staggered chemical potential $\mu_{i}=\mu_{0} e^{iQr_i}$ and/or a staggered magnetic field $h_{i}=h_{0} e^{iQr_i}$, with ($Q=(\pi,\pi)$ for both AFM and checkerboard CO) :
\begin{equation}
H_{\mu_0,h_0}=H+\sum_{i\sigma} \mu_{i} n_{i\sigma}+\sum_{i} h_{i}m_{i},
\label{Eq.2}
\end{equation}
here 
%and $\mu_i=\mu_0e^{iQr_i}$,$h_i=h_0e^{iQr_1}$, 
 $m_i=n_{i,\uparrow}-n_{i,\downarrow}$. The staggered chemical potential and magnetic field break the translational symmetry and divide the original bipartite lattice into two sub-lattices $A$ and $B$, 
%$\mu_{i\sigma}=\pm \mu_{0\sigma}$ and $h_{i\sigma}=\pm h_{0\sigma}$ for sub-lattice $A(B)$ respectively,
thereby doubling the unit cell. 
%AFM is described by $h_{0\uparrow} = -h_{0\downarrow}$, while charge order is described by $\mu_{0\uparrow} = \mu_{0\downarrow}$.  
In this paper, we begin simulations with a small $\mu_{0}/t \approx 0.05$ or  $h_{0}/t \approx 0.05$ on the first iteration and then set these quantities to zero on subsequent iterations.  The system is then allowed to evolve freely, and will either converge to a paramagnetic state (electrons uniformly distributed over lattice site and spin) or fall into one of the ordered states.  

Solving the cluster impurity problem requires the use of a quantum impurity solver. Here we use the continuous time auxiliary field quantum Monte Carlo algorithm (CTAUX)\cite{Gull08,Gull11,Gull11_RMP}, modified to accommodate non-local density-density interactions \cite{Terletska2017, Terletska2018}.

\subsection{Green's Functions}
The ordered phases investigated here reduce the translation symmetry of the lattice.\cite{Maier2005}  This doubles the size of the unit cell in real space while halving the size of the Brillouin zone, such that in the ordered phase the momentum space points $k$ and $k+Q$ become degenerate, where for AFM and CO $Q=(\pi,\pi)$.  In order to study ordered and non-ordered phases with the same method, a double cell formalism is used in which momentum space Green's functions take on a block diagonal structure.  Each block takes on the form
\begin{align}
    G_\sigma(k,i\omega_n) = \left( \begin{array}{cc}
    G_\sigma(k,k; i\omega_n) & G_\sigma(k,k+Q; i\omega_n)  \\
    G_\sigma(k+Q,k; i\omega_n)  &  G_\sigma(k+Q,k+Q; i\omega_n)
    \end{array} \right),
\end{align}
where $\omega_n = (2n +1)\pi/\beta$ with $\beta=1/k_BT$ denotes the fermionic Matsubara frequencies. In the absence of order $G_\sigma(k,k+Q)=G_\sigma(k+Q,k)=0$ ~\cite{Terletska2017, Fuchs2011}, so that the Green's functions become diagonal in momentum space. In the ordered phases these off diagonal components become finite and obey symmetry relations.  For AFM, $G_\sigma(k,k+Q)=G_{\sigma}(k+Q,k)=-G^*_{-\sigma}(k,k+Q)=-G^*_{-\sigma}(k+Q,k)$, while for CO we have $G_\sigma(k,k+Q)=G_{\sigma}(k+Q,k)=G_{-\sigma}(k,k+Q)=G_{-\sigma}(k+Q,k)$.
Here on, for a short-hand notation we drop the frequency index.
We can define both the momentum dependent and local sublattice and spin resolved Green's functions as follows. \cite{Terletska2017}

\begin{align}
G_{A/B,\sigma}(k) &= \frac{G_\sigma(k,k)+G_\sigma(k+Q,k+Q)}{2} \pm G_\sigma(k,k+Q)
\label{Eq:Gk} \\
G_{A/B,\sigma}^{loc} &= \frac{1}{N_C}\sum_k G_{A/B,\sigma}(k)
\label{Eq:Gloc} 
\end{align} 

Similar equations describe the sublattice resolved self-energies.  These quantities allow us to study how the density of states (from analytic continuation of $G_{A/B,\sigma}^{loc}$) and self-energies behave on each sublattice.

\subsection{Order Parameter}
The order parameters for charge order, $\Delta_{CO}$, and anti-ferromagnetism, $\Delta_{AFM}$, can be computed from the spin resolved cluster site densities, $n_{i\sigma}$.

\begin{align}
\Delta_{CO} &= \frac{2}{N_c}\left | \sum_{i\in A, \sigma}n_{i\sigma} - \sum_{i\in B, \sigma}n_{i\sigma}\right |
\label{Eq:DelCO}\\
\Delta_{AFM}&=\frac{1}{N_c}\sum_i \left |  n_{i\uparrow}- n_{i\downarrow} \right |
\label{Eq:DelAFM}
\end{align}

These expression can also be written in terms of the off diagonal components of the momentum space Green's function in imaginary time, $G_{k,k+Q,\sigma}(\tau)$.\cite{Maier2005}

\begin{align}
\Delta_{CO} &= \frac{2}{N_c}\left |\sum_{k\sigma} G_{\sigma}(k,k+Q;\tau=0^-) \right |
\label{Eq:DelCO_G}\\
\Delta_{AFM}&=\frac{1}{N_c}\left |\sum_{k}\left(G_{\uparrow}(k,k+Q;\tau=0^-) - G_{\downarrow}(k,k+Q;\tau=0^-) \right)\right |
\label{Eq:DelAFM_G}
\end{align}

%%%%%%%%%%%%%%%%%%%%%%%%%%%%%
\section{Results}\label{sec:Results}
%%%%%%%%%%%%%%%%%%%%%%%%%%%%%%
We present the dependence of phase boundaries, order parameters, and of the energetics on $U$, $V$, and $T$ for the half-filled extended Hubbard model.  We focus on  three  phase boundaries exhibited by the model in our approximation: those between the non-ordered paramagnetic (normal) and antiferromagnetic phases (Normal-AFM), normal and charge ordered phases (Normal-CO), and antiferromagnetic and charge ordered phases (AFM-CO).  In section \ref{ssec:TVpd}, we present the $T-V$ phase diagram (at $U=4t$) for the model and examine how the order parameters and energetics behave along cuts through the different phase boundaries.  We also demonstrate hysteresis across the AFM-CO and Normal-CO phase boundaries, indicating  first order transitions.  In section \ref{ssec:UVpd}, we present the temperature dependence of the $V-U$ phase diagram, comparing temperatures $T/t=1/6$ and $T/t=1/10$ .  

\begin{figure}[t!]
  \centering
    \includegraphics[width=0.45\textwidth]{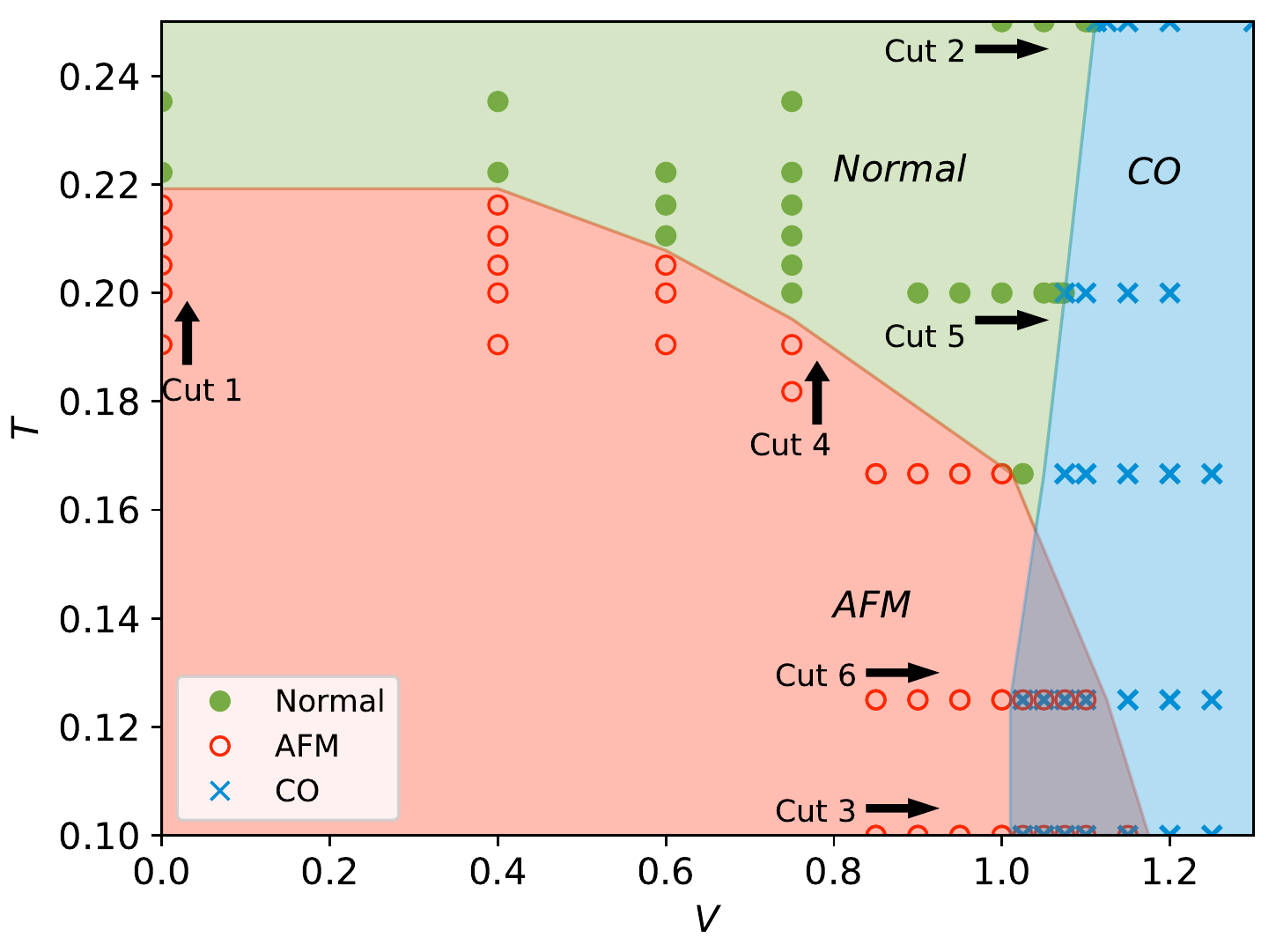}
    \caption{$T-V$ phase diagram for the half-filled extended Hubbard model at $U=4$. Green shading and filled circles represent normal (paramagnetic) state. Red area and open circles represent region with AFM ordering. Blue area and crosses depict the CO state. Region with both crosses and circles shows the first order CO/AFM coexistence. Symbols denote simulation points. Transition lines are obtained from the midpoint between simulation points.
    Also indicated are six phase transition cuts referred to in the text.
  }\label{fig:TV_Phase_Diagram}
\end{figure}

\subsection{$T$-$V$ phase diagram}\label{ssec:TVpd}
The phase diagram as a function of nearest neighbor interaction $V$ and temperature $T$ at fixed $U/t=4$ is shown in Fig. ~\ref{fig:TV_Phase_Diagram}. The model exhibits a paramagnetic metallic phase (from now on referred to as the normal state) at high temperature and weak $V$ (green shading/filled circles in Fig.~\ref{fig:TV_Phase_Diagram}), an AFM (red shading/open circles) at low temperature and low $V$, and a CO state at large $V$ (blue shading/crosses). Symbols indicate simulation points; the phase transition boundaries  are obtained from the midpoint between simulation results in different phases. 

As is expected from Hubbard model simulations in the absence of $V$, strong AFM correlations exist at half filling. In cluster DMFT simulations, these cause the system to polarize and fall into a long-range antiferromagnetically ordered phase \cite{Lichtenstein00,Maier2005} below a transition temperature of $T \sim 0.22$ (at $V=0$). This `phase' is an artifact of the approximation and should be understood as an area where long-ranged AFM fluctuations are strong.~\cite{Maier2005}

Larger DCA clusters will eventually lead to a suppression of  AFM order in 2D and simply exhibit strong AFM fluctuations \cite{Maier2005}. The AFM correlation length is large compared to accessible cluster sizes (and rapidly growing as temperature is decreased), making observing a true paramagnetic state difficult  within this approximation.\cite{Maier2005} However, one may expect that effects present in real systems but excluded from the Hubbard model, such as inter-layer couplings, may stabilize these fluctuations and lead to an actual phase transition with similar overall behavior.

Non-local interactions $V$ suppress these fluctuations. We find that as we increase $V$ above $\sim 0.6t$, the critical temperature of the AFM phase is rapidly reduced. Within DCA, further increase of $V$ will entirely suppress the AFM state, so that beyond a value of $\sim 1.2t$ no AFM ordering is observed in our calculations.

Repulsive non-local interactions on a bipartite lattice eventually lead to a charge ordered state \cite{Terletska2017}. For our parameters, at $U/t=4$, this charge ordering sets in at $V/t \sim 1.1$ for the highest T shown. Lowering the temperature shifts that phase boundary towards lower values of $U$, such that at $T/t=0.1$ the phase boundary is observed near $V/t=1.0$.

For the parameter values chosen, there is an area where both CO and AFM states can occur. In this area, the nearest neighbor interaction $V$ is large enough that CO is favorable, but the temperature is low enough that AFM fluctuations are strong. In our simulation, we find a first-order coexistence regime where the system is either in a CO state (where magnetic order is absent) or in an AFM state (where charge order is absent).

\begin{figure*}[tbh]
  \centering
    \includegraphics[width=0.95\textwidth]{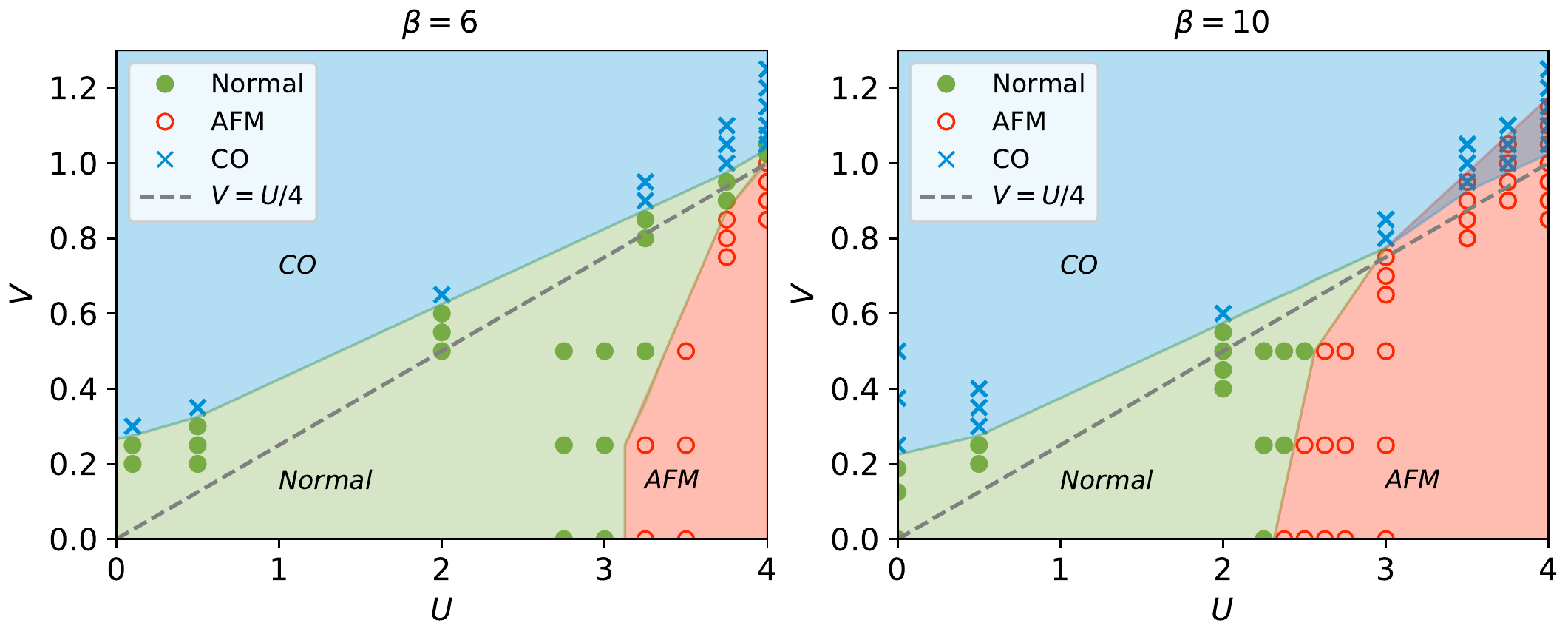}
    \caption{
    Comparison of the $U-V$ phase diagram of the extended Hubbard model at inverse temperature $\beta=6/t$ and $\beta=10/t$, both at $\mu=0$.  At high temperature (left panel) the AFM and CO phases are entirely separated by the normal state region for this range of $U$.  Upon lowering $T$, a hysteresis region emerges at larger $U$ in which both the CO and AFM solutions are stable. Points with both blue crosses and red open circles indicate points at which the simulation converges to either a CO or AFM solution, depending on whether a CO or AFM starting solution is used.  The mean field result (dashed line) for the phase boundary between the normal and CO state, $V=U/4$, is also shown.}\label{fig:UV_PD}
\end{figure*}
\subsection{$V$-$U$ phase diagram}\label{ssec:UVpd}
To illustrate the evolution of the phase diagram as a function of  $U$ and $V$, we present cuts in the $V$-$U$ plane at constant temperature in Fig.~\ref{fig:UV_PD}. The left panel shows $T/t=1/6$, the right panel $T/t=1/10$. At large $V$, the system is charge ordered at half-filling (blue area). At small $V$ but large $U$, the system undergoes an AFM transition in this approximation (red area). And at small $U$ and $V$, the model is in an isotropic `normal' state (green). 

As explored in previous work \cite{Terletska2017,Terletska2018} (see also results from other methods \cite{Merino01, Wolff83,Yan1993,vanLoon, silke_Fock}), the CO transition occurs above the mean field line \cite{Bari71}, has a non-zero intercept at $U=0$, and is only weakly temperature dependent. In contrast, the AFM phase in this approximation is very strongly temperature dependent for these parameters, hinting at a rapid evolution of the spin susceptibility in this model, and moves to substantially lower $U$ and larger $V$ as the temperature is lowered (compare to the right panel).

At the lower temperature, the coexistence between the two phases occurs at large $V$ and large $U$, where the non-local interaction is strong enough to favor CO but the local $U$ also permits long range AFM.

\begin{figure}[tbh]
  \centering
    \includegraphics[width=0.45\textwidth]{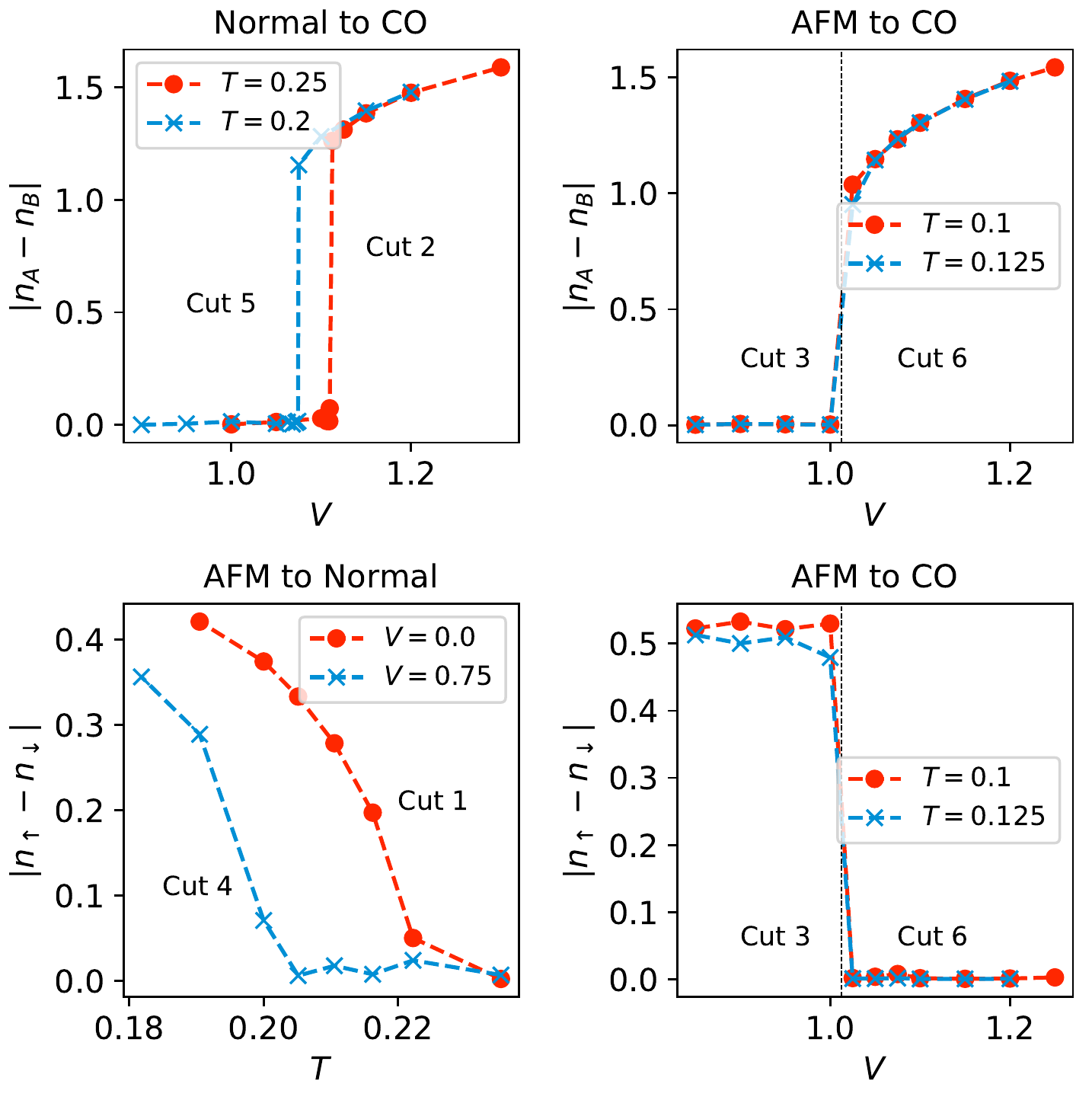}
    \caption{AFM and CO order parameters across phase transition. $U/t=4$, $\mu=0$.  Normal-CO at constant $T$ (top left panel), AFM-Normal at constant $V$ (bottom left panel), and AFM-CO (right panels).  AFM-CO cuts are obtained with a CO starting solution; see Fig.~\ref{fig:AFM_CO_hyst} for hysteresis.}\label{fig:OrderParams}
\end{figure}

\begin{figure*}[tbh]
  \centering
    \includegraphics[width=0.95\textwidth]{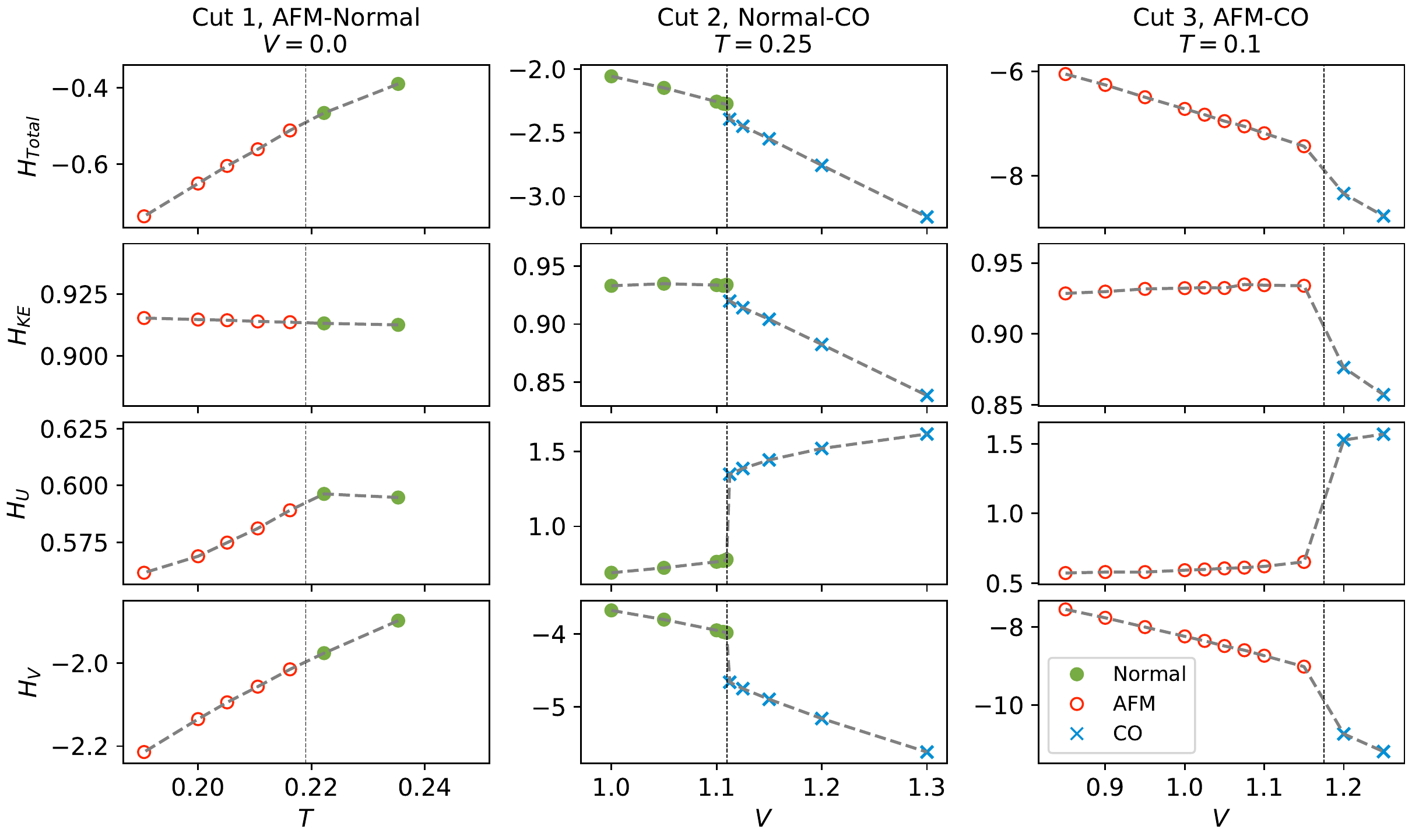}
    \caption{Contribution to the energetics across three phase transitions for the extended Hubbard model at $U=4$, $\mu=0$. Left Column: AFM-Normal transition along Cut~1 from Fig. \ref{fig:TV_Phase_Diagram}.  As AFM order emerges at lowering temperature, the on site interaction energy, $H_U$ is suppressed by the reduction in double occupancy.  Middle Column: Normal-CO transition along Cut~2 from Fig. \ref{fig:TV_Phase_Diagram}. The localization of electrons on the one sublattice leads to a decrease in the kinetic energy, $H_{KE}$, an increase in the on-site interaction energy, $H_U$, in exchange for a decrease in the nearest neighbor interaction energy, $H_V$. Right Column: AFM-CO transition along cut 3 from Fig. \ref{fig:TV_Phase_Diagram}, showing only the results obtained from the AFM starting solution.  An increase in the on-site interaction energy, $H_U$ is exchanged for a decrease in the nearest neighbor interaction energy, $H_V$. The symbols for each data point, indicating the stable phase, follow from Fig.~\ref{fig:TV_Phase_Diagram}.  }\label{fig:EnergyCuts}
\end{figure*}

\subsection{Order parameter and phase boundaries}\label {ssec:order_parameter}
CO is characterized by a difference between the occupancies on different sublattices, as described by the order parameter in Eq.~\ref{Eq:DelCO}. AFM, as defined by the order parameter of Eq.~\ref{Eq:DelAFM}, is identified by different occupancies of the two spin species. In order to distinguish between ordered and isotropic points in the presence of Monte Carlo noise, we define simulation points with order parameters larger than $0.1$ as ordered in Figs.~\ref{fig:TV_Phase_Diagram} and \ref{fig:UV_PD}.

Raw data for the order parameters along the cuts indicated in Fig.~\ref{fig:TV_Phase_Diagram} are shown in Fig.~\ref{fig:OrderParams}. The bottom left panel shows the order parameters across the AFM to Normal state phase boundary. Shown are two cuts at constant $V$ but for varying temperature. As expected, this phase transition is second order \cite{Terletska2017}. Larger non-local interaction $V=0.75t$ moves the onset of the phase transition to lower temperatures, suppressing both the onset and the strength of the AFM order parameter. 

The top left panel shows the transition from the normal state to CO, at constant temperature $T$, as a function of $V$. CO phase is identified by a non-zero staggered density appearing at larger values of $V$. In the absence of long-ranged AFM order, this transition has been analyzed in detail in previous work \cite{Terletska2017}. As discussed later on in (Sec.~\ref{ssec:hysteresis}), at larger values of local interactions $U$, we find this transition to be the first order transition~\cite{Potthoff,Schuler_2019,Schuler18} with a characteristic hysteresis behavior of the order parameter. Lower temperatures lead to an earlier onset of the CO state at lower $V$.\cite{Terletska2017}

The right two panels show the order parameter across the transition from the AFM (low $V$) to the CO (large $V$) states, at constant $T$ as a function of $V$. Shown are both the magnetic (bottom) and CO (top) order parameters. This transition is first order (see Sec.~\ref{ssec:hysteresis} for hysteresis); shown here are data obtained by starting from a CO solution.

\subsection{Energetics}\label{ssec: Energetics}
Fig.~\ref{fig:EnergyCuts} shows the contributions to the energetics as the system crosses the phase boundaries. Shown are the total energy $H_{Total}$, the kinetic energy $H_{KE}$, the contribution of the local energy to the interaction energy $H_U$, and the contribution of the non-local term to the interaction energy $H_V$. These energies are computed as~\cite{Gull11_RMP,Haule07}
\begin{align}
H_{KE} &= \frac{1}{N_C}\sum_{k\sigma} (\epsilon_k-\tilde{\mu})\langle n_{k\sigma} \rangle
\label{Eq:HKE} \\
H_{V} &=  \frac{K-\langle k \rangle}{\beta N_C} - H_U
\label{Eq:HV} \\
H_{U} &= \frac{U}{N_C}\sum_i \langle n_{i\uparrow}n_{i\downarrow}\rangle,
\label{Eq:HU}
\end{align}
and $H_{Total} = H_{KE}+H_U+H_V$. 
$\epsilon_k=-2t(\cos k_x+\cos k_y)$ is the dispersion on 2D square lattice, $\langle k \rangle$ denotes the average order sampled during the Monte Carlo simulation\cite{Gull11_RMP,Haule07} and $K$ is a constant introduced in the CTAUX algorithm by a Hubbard-Stratonovich transformation.\cite{Gull08}

The first column of Fig.~\ref{fig:EnergyCuts} shows how the different energy components change as the temperature is increased and the system moves from an AFM ordered phase to the normal state.  The dominant change upon entering the AFM phase is a reduction of the on-site interaction, $H_U$, due to the suppression of the double occupancy. Kinetic energies and potential energies show little change across the transition. This is consistent with the AFM transition in single site DMFT and four-site cluster DMFT below the Mott transition, where the opening of the AFM gap lowers the energy by suppressing the double occupancy.\cite{Gull08_plaquette}

The second column shows the energetics as $V$ is increased and the system enters the CO state from the normal state at high temperature ($T=0.25$).  Here, the major change in the energetics is the non-local interaction energy term $H_V$, which can be dramatically lowered by entering a CO phase. The kinetic energy decreases slightly as electrons become constrained to one sublattice, and the transition is accompanied by an increase in the on-site interaction energy, $H_U$, caused by the increase of the double occupancy in the CO state.  The sharp jump in energies across the transition is consistent with the first order transition across this cut. 

Finally, the third column displays the evolution of the system across AFM-CO transition at lower temperature $T=0.1$. The transition is the first order with a very pronounced jump in energy changes across the phase boundaries. The data shown is from the branch of the hysteresis that starts in the AFM phase. It is evident that the transition requires a substantial rearrangement of the energetics, with major changes in all energy terms.

\begin{figure}[tbh]
  \centering
    \includegraphics[width=0.45\textwidth]{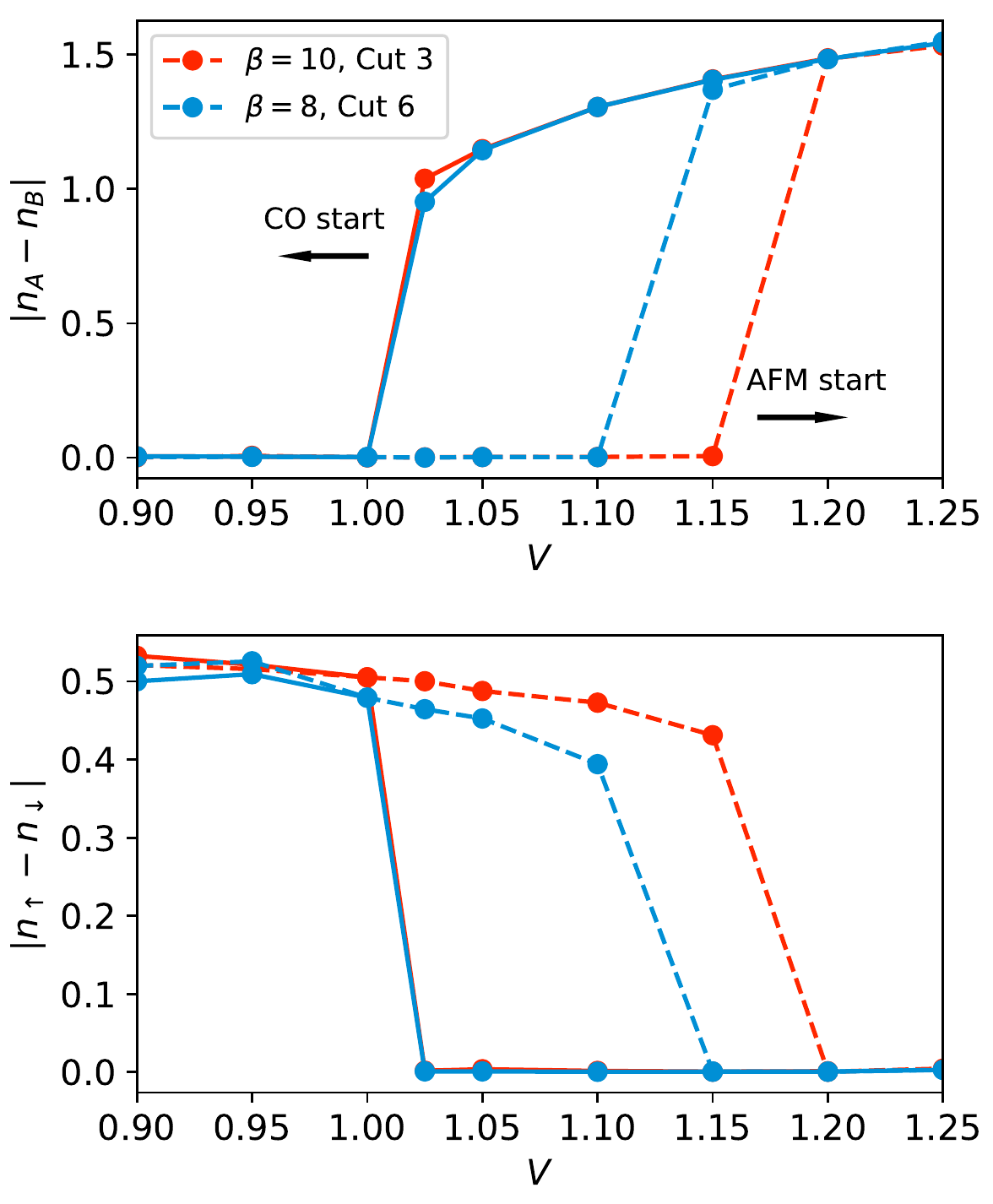}
    \caption{Hysteresis between AFM and CO state. $U=4t$, $\beta t=10$ (red) and $\beta t=8$ (blue). Top panel: CO order parameter, $\Delta_{CO}$. Bottom panel: AFM order parameter, $\Delta_{AFM}$. Dashed (solid) lines indicate convergence from a CO (AFM) initial guess.}
    \label{fig:AFM_CO_hyst}
\end{figure}

\begin{figure}[tbh]
  \centering
    \includegraphics[width=0.45\textwidth]{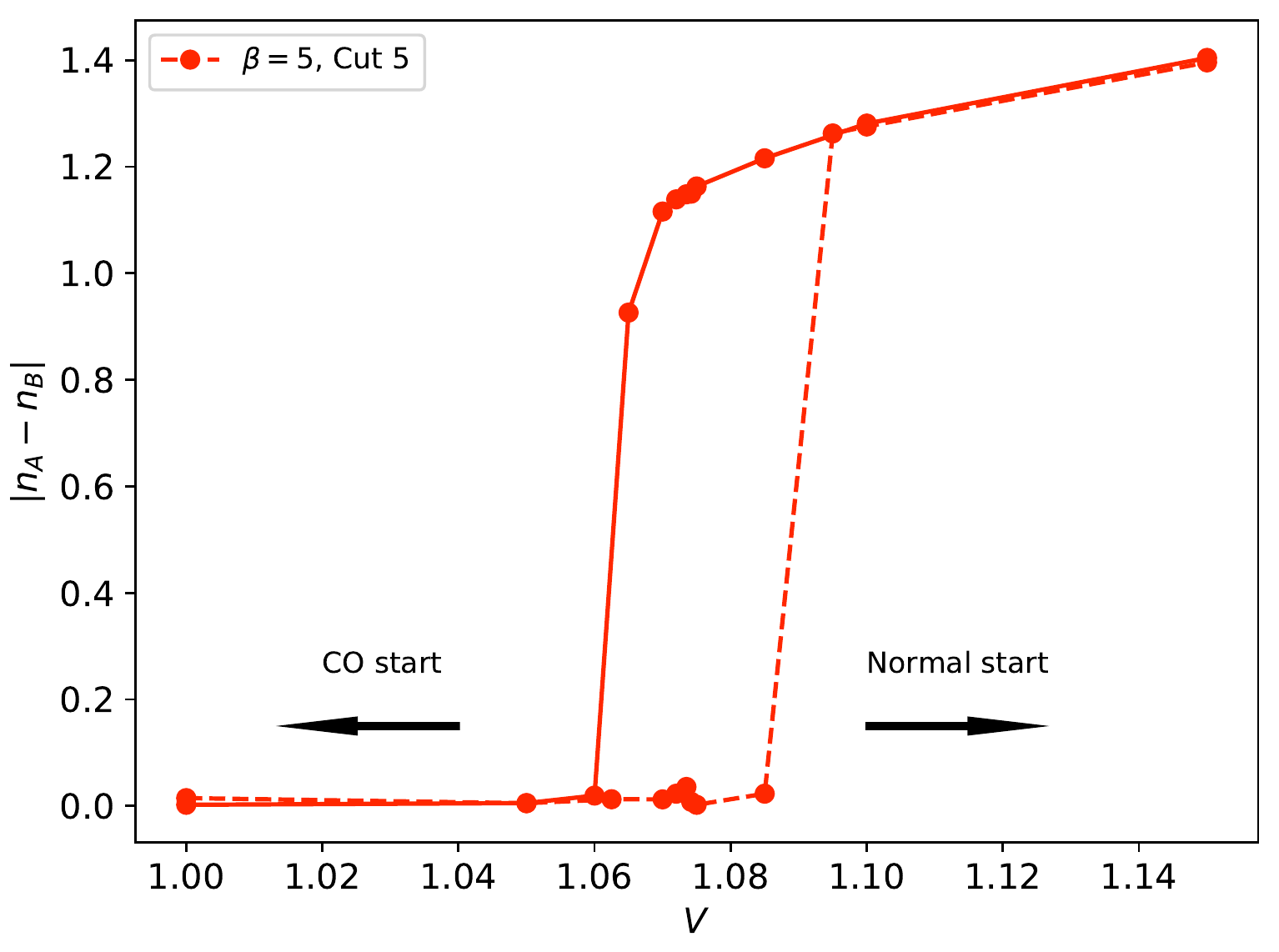}
    \caption{Hysteresis between Normal and CO states. $U=4t$, $\beta t=5$. The data indicate the hysteresis behavior across the transition depending on the starting solution. Shown is the converged CO order parameter, $\Delta_{CO}$, arising from simulations started with Normal solution with a small CO offset (dashed) and a CO solution (solid). }\label{fig:CO_Normal_hyst}
\end{figure}

\begin{figure*}[tbh]
  \centering
    \includegraphics[width=0.95\textwidth]{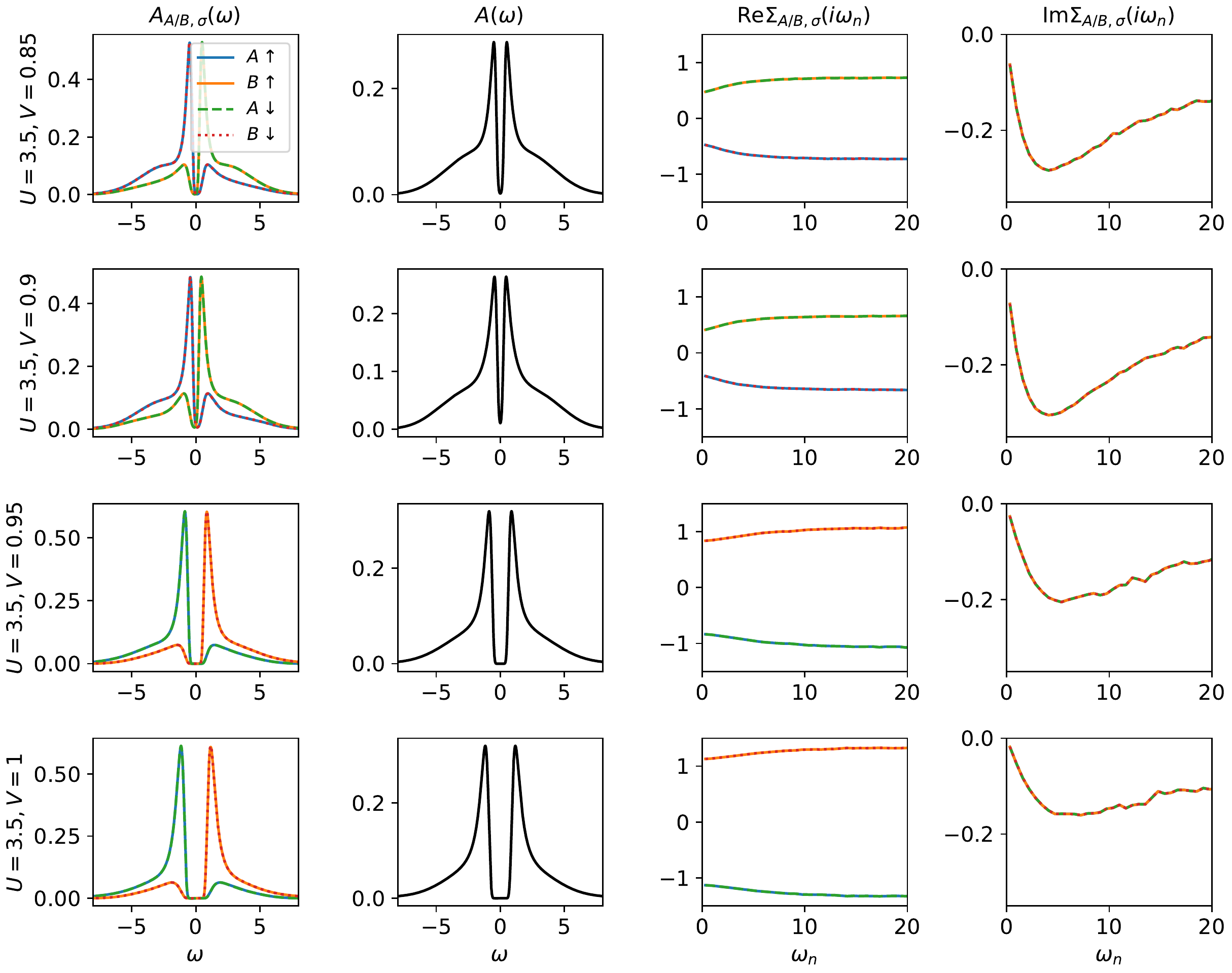}
    \caption{Evolution of spectral functions across AFM-CO phase boundary at $U=3.5t$.  First column: spin and sublattice resolved spectral function. Second column: local spectral function depicting the qualitative difference between the small AFM and large CO gap. Third and fourth columns: real and imaginary part of the Matsubara frequency self-energy.}\label{fig:DoS_SE_Normal_AFM_CO}
\end{figure*}

\subsection{Hysteresis} \label{ssec:hysteresis}

We present evidence of hysteresis in the AFM/CO transition at low temperatures in Fig.~\ref{fig:AFM_CO_hyst}.  This data is obtained by running each simulation point twice - once with an initial configuration corresponding to an AFM ordered state, and once with one describing a CO state. Outside the coexistence region both of these simulations converge to the same solution.  In contrast, in a coexistence region both states will be stable and the two simulations will converge to different solutions.

The top panel of Fig.~\ref{fig:AFM_CO_hyst} shows the CO order parameter, while the bottom panel shows the AFM order parameter. Shown is a trace along Cut 3 ($\beta t=10$) and Cut 6 ($\beta t=8$) as a function of $V$. A coexistence regime starts at $V/t\sim 1$ and extends to $V/t\sim 1.2$ at the lower temperature, and shrinks as temperature increases (demonstrated by the $T/t=1/8$ data) and eventually vanishes, see Fig.~\ref{fig:TV_Phase_Diagram}. The data indicates that the stable states are always only AFM or CO, and that no solutions have both finite AFM and finite CO ordering.

We also find the evidence for a small hysteresis region in the Normal/CO transition shown in Fig.~\ref{fig:CO_Normal_hyst}.  The figure shows the converged CO order parameter resulting from two sets of simulations, at $\beta t=5$ and $U/t=4$.  In the first set, each simulation is started with a Normal state solution with a small CO offset.  In the second set, each simulation is started with a CO state solution.  For a narrow range of $V$, from about $V/t \approx1.065$ to $V/t \approx 1.09$, these simulations reveal that both Normal and CO states are stable.  This indicates that at this temperature and interaction strength, the Normal to Charge Order phase transition is first order. 

This finding is consistent with the sharp transition in energy displayed in Fig.~\ref{fig:EnergyCuts}, as well as previous work\cite{Terletska2017} that indicated that the Normal to CO transition is continuous at small $U$ but sharpens as $U$ is increased\cite{Potthoff, Schuler18, Schuler_2019}.  Since the hysteresis region is so narrow, we do not attempt to draw it on our phase diagrams.  All other plots in this paper dealing with the Normal to CO transition display data obtained from simulations that start with a Normal state solution.

\subsection{Spectral Functions}\label{ssec:spectral_function}
 Fig.~\ref{fig:DoS_SE_Normal_AFM_CO} shows the evolution of the spectral function and self-energy across the AFM/CO phase boundary at $U=3.5t$ and $\beta t=10$.  The first column depicts the different sublattice and spin contributions to the total spectral function, which is shown in the second column.  The symmetry of the sublattice and spin components vary as described in Sec.~\ref{sec:Methods}.  At lower $V$ the occupied states (i.e. states with energy below $\omega=0$) are predominantly those with spin up on the A sublattice and spin down on the B sublattice. These components are equal to each other and related to the other components (spin down on sublattice A and spin up on sublattice B) by particle-hole symmetry (i.e. $\omega \rightarrow -\omega$), as expected for an AFM state. Upon increasing $V$ and transitioning into the CO state, the symmetry of the spectral function components change so electrons occupy the A sublattice and vacate the B sublattice, with symmetry between the up and down spin components.

We can use the total spectral function results to compare the energy gaps at $\omega=0$ on either side of the AFM/CO transition.  At lower $V$, the system is not fully gapped and is in a metallic state with AFM order. In contrast, the CO state displays a gap immediately after the transition.

The last two columns of Fig.~\ref{fig:DoS_SE_Normal_AFM_CO} show how the real and imaginary parts of the sublattice and spin resolved Matsubara self-energies behave through the AFM/CO transition.  The real parts switch symmetry and increase in magnitude upon entering the CO state, in agreement with the formation of a robust electronic gap.  In contrast, the imaginary part of the self-energy seems to be smaller in the charge order state than the AFM state, indicating smaller correlation effects.  This behavior makes physical sense because the AFM state is dependent upon spin correlations between electrons in the two sublattices (i.e. virtual exchange hopping), whereas the CO state can be viewed as a result of classical energetics that favor a reduction in the double occupancy.

Spectral functions were obtained via the Maximum Entropy Method (MEM), as implemented by the ALPS MaxEnt software package. \cite{Levy17,Wallerberger18}
%%%%%%%%%%%%%%%%%%%%%%%%%%%%%
\section{Conclusion}\label{sec:Conclusions}
%%%%%%%%%%%%%%%%%%%%%%%%%%%%%%
%In this paper we present a detailed discussion of the competition between the normal, AFM, and CO states in the 2D half-filled extended Hubbard model on a square lattice.  The calculations are performed with the DCA method extended to ordered phases on an 8-site cluster, and a modified version of the CTAUX impurity solver was used to include non-local interactions.

%Our main results include phase diagrams for the model that indicate how nearest neighbor interactions $V$ suppress AFM fluctuations and lower the critical temperature below which the AFM phase exists in this approximation. We also detail the energetics of the three different transitions (Normal to AFM, AFM to CO, and Normal to CO).

%Our work reveals the existence of hysteresis in both the Normal to CO and CO to AFM phase transitions.  These results suggest that the CO to AFM phase transition is always first order, while the Normal to CO transition may only be first order at larger interaction strengths.  Characterizing the Normal to CO transition across interaction strengths may be an interesting topic for further study.

%Finally, spectral functions across the AFM to CO phase transition are presented to emphasize the underlying symmetries and illustrate the changes in the electronic gap.

In conclusion, we have presented results for the two-dimensional half-filled extended Hubbard model. Within the DCA approximation, the model exhibits both AFM and CO originating from strong electronic correlations. These orders are stable in a large part of parameter space, allowing us to probe the behavior of physical observable in the vicinity of the phase transitions as well as deep within a phase.

We find that the non-local interactions, which promote screening and CO, also strongly suppress AFM. Nevertheless, there is a phase coexistence regime. Phase boundaries are consistent with a continuous transition in the case of the Normal-AFM transition, and are first order (we show hysteresis) in case of the Normal-CO and AFM-CO boundary. A detailed analysis of the energetics, of order parameters, and of the spectral functions is provided.

Real materials that exhibit CO in the vicinity of AFM are considerably more complex than the simple extended Hubbard model. Nevertheless, there is merit in identifying model systems and non-perturbative approximations in which those phases occur in close proximity, as simple competition effects such as the one between local and non-local interactions here can provide general guiding principles for understanding their overall behavior.

While the exact solution of the two-dimensional model does not support long-range ordered AFM, none of the physical compounds are idealized two-dimensional systems. The role of a weak inter-layer coupling or of other band structure effects is mimicked by the short-ranged nature of the DCA approximation.

It would be very interesting to examine if other ordered phases, such as superconductivity, emerge in the vicinity of AFM and CO. The temperatures accessible in our systems are much too high to address this question directly, though other techniques such as a susceptibility analysis\cite{Chen15,Toschi15} may be employed. We therefore leave this question open for further study.

\acknowledgments{
This work was supported by NSF DMR 1606348. Computer time was provided by NSF XSEDE under allocation TG-DMR130036.
}

%\clearpage
%\newpage
%\pagebreak

\bibliographystyle{apsrev4-1}
\bibliography{CO_refs.bib}
\end{document}